\documentclass[10pt,letterpaper]{article}

\usepackage{cogsci}

\cogscifinalcopy 

\usepackage{pslatex}
\usepackage{float} 
\usepackage{cuted} 
\usepackage{etoolbox} 
\usepackage{amsmath} 
\usepackage{graphicx} 
\usepackage{hyperref} 
\usepackage{apacite} 

\title{A Neural Network Model of Continual Learning with Cognitive Control}
 
\author{{\large \bf Jacob Russin$^{1,4}$ (jlrussin@ucdavis.edu)} \\
  \And {\large \bf Maryam Zolfaghar$^{2,4}$ (mzolfaghar@ucdavis.edu)} \\
  \AND {\large \bf Seongmin A. Park$^3$ (apark@ucdavis.edu)} \\
  \And {\large \bf Erie Boorman$^{1,3}$ (edboorman@ucdavis.edu)} \\
  \AND {\large \bf Randall C. O'Reilly$^{1,2,4}$ (oreilly@ucdavis.edu)} \\
  }

\begin{document}

\setlength{\abovedisplayskip}{3pt}
\setlength{\belowdisplayskip}{3pt}

\maketitle
\begin{strip}
{\centering
    $^1$ Department of Psychology, UC Davis; $^2$ Department of Computer Science, UC Davis;
    
    $^3$ Center for Mind and Brain, UC Davis;  $^4$ Center for Neuroscience, UC Davis
\par}
\end{strip}

\begin{abstract}
Neural networks struggle in continual learning settings from catastrophic forgetting: when trials are blocked, new learning can overwrite the learning from previous blocks. Humans learn effectively in these settings, in some cases even showing an advantage of blocking, suggesting the brain contains mechanisms to overcome this problem. Here, we build on previous work and show that neural networks equipped with a mechanism for cognitive control do not exhibit catastrophic forgetting when trials are blocked. We further show an advantage of blocking over interleaving when there is a bias for active maintenance in the control signal, implying a tradeoff between maintenance and the strength of control. Analyses of map-like representations learned by the networks provided additional insights into these mechanisms. Our work highlights the potential of cognitive control to aid continual learning in neural networks, and offers an explanation for the advantage of blocking that has been observed in humans.

\end{abstract}

\section{Introduction}
Neural networks have shown impressive performance on many problem domains in machine learning (ML), where they are typically trained on batches of data that are independent and identically distributed \cite{HadsellRaoRusuEtAl20}.
However, agents learning about the world in real time experience streams of data that are not independent (e.g., a human may spend one day exploring one part of an unfamiliar city, and spend the next day exploring another part).
The neural networks that have driven recent success in artificial intelligence perform poorly in these continual-learning settings because of the well known phenomenon of catastrophic forgetting \cite<or catastrophic interference;>{McClellandMcNaughtonOReilly95, McCloskeyCohen89}.
When samples or trials are blocked, learning in new blocks overwrites the learning that occurred in previous blocks.
Humans and other animals do not exhibit such extreme forgetting \cite{McClellandMcNaughtonOReilly95}, and in some cases even demonstrate an \textit{advantage} when trials are blocked \cite{CarvalhoGoldstone14, FleschBalaguerDekkerEtAl18a, NohYanBjorkEtAl16, WulfShea02}.
This suggests that there are mechanisms in the brain that mitigate catastrophic forgetting and can even reverse this effect, making it easier to learn when experiences are correlated over time.

A number of strategies for overcoming catastrophic forgetting in neural networks have been proposed in both computational neuroscience \cite{FleschBalaguerDekkerEtAl18a, McClellandMcNaughtonOReilly95} and ML \cite{BotvinickRitterWangEtAl19, HadsellRaoRusuEtAl20, MnihKavukcuogluSilverEtAl13, VelezClune17a}. 
Complementary learning systems (CLS) theory emphasizes that catastrophic forgetting arises when learning occurs too quickly in overlapping representations \cite{McClellandMcNaughtonOReilly95, OReillyBhattacharyyaHowardEtAl11}, and that the episodic memory system in the hippocampus plays an important role in learning representations that are sparse or pattern-separated, allowing rapid learning to take place. 
These ideas have inspired ML researchers to deploy memory systems that replay past experiences in a relatively independent fashion, thereby overcoming catastrophic forgetting \cite{BotvinickRitterWangEtAl19, MnihKavukcuogluSilverEtAl13}.

Constraining patterns of activity to be sparse is not the only way to ensure they will not overlap and interfere with each other.
Theories of cognitive control in the prefrontal cortex (PFC) emphasize that a crucial function of control is to selectively modulate activity in other brain areas in order to coordinate a response that aligns with the current context or goal \cite{HerdOReillyHazyEtAl14, MillerCohen01, RougierNoelleBraverEtAl05}. 
Cognitive control may therefore play an important role in regulating learning in other brain regions so that patterns of activity do not overlap across different contexts or goals \cite{RougierNoelleBraverEtAl05, TsudaTyeSiegelmannEtAl20}. 

Here, we build on this work and test neural networks in conditions where trials are either blocked or interleaved, showing how cognitive control can help to mitigate catastrophic forgetting in the blocked condition. 
We further hypothesized that in some cases learning across blocked trials is \textit{superior} to interleaving because of an internal bias of the PFC to maintain its activity over time, creating a cost to rapidly switching between contexts or goals \cite{BlackwellChathamWiseheartEtAl14, HerdOReillyHazyEtAl14, OReillyFrank06}. 
This idea fits well with a general framework where the cost of switching must be traded off against the strength of control: stronger control results in less catastrophic forgetting, but more difficulty switching \cite{HerdBanichOReilly06, ShenhavBotvinickCohen13a}.
We perform our simulations on a task designed to induce learning of map-like representations \cite{ParkMillerBoorman21a, ParkMillerNiliEtAl20, RussinZolfagharParkEtAl21} so that we could additionally investigate how cognitive control affected the model’s representations.

\subsection{Task}

\begin{figure}[ht]
\begin{center}
  \includegraphics[width=0.5\textwidth]{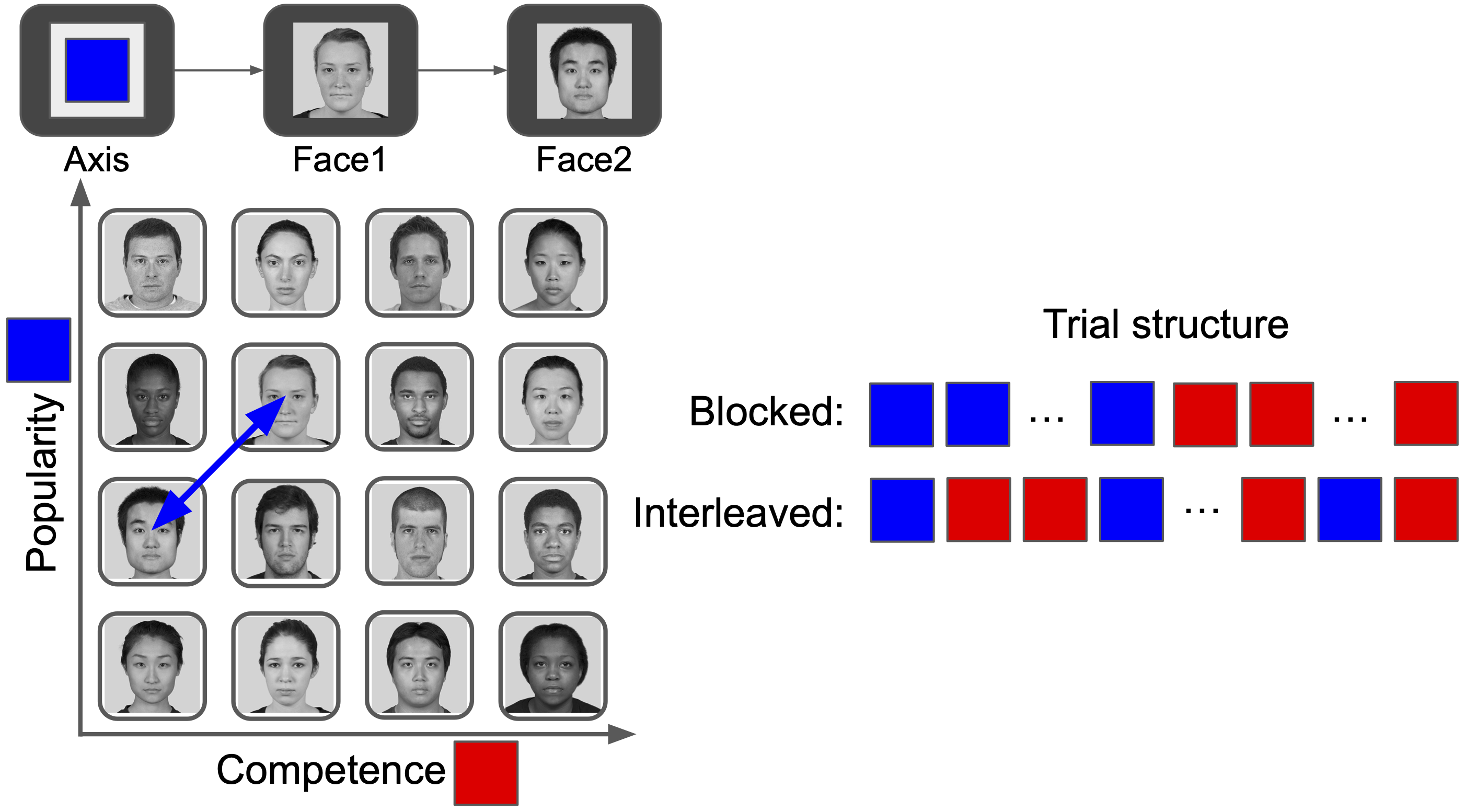}
  \caption{Task structure. The model learned the relative ranks of people along two social hierarchy dimensions: popularity and competence. The model learned through trial and error to select which of two faces ranked higher along one of the two dimensions (indicated by a cue). Trials were either interleaved, where cues were randomly shuffled, or blocked, where one dimension was learned at a time.}
\label{fig:task}
\end{center}
\end{figure}

We trained neural network models on an existing task taken from an fMRI experiment \cite{ParkMillerNiliEtAl20, RussinZolfagharParkEtAl21}.
Participants in the experiment learned about the relative ranks of 16 people in a hypothetical social hierarchy along two separate social dimensions (see Figure \ref{fig:task}). 
In the instructions of the task, these dimensions were presented as ranks in “popularity” and “competence.” 
On each trial, the participants were shown two faces, and were cued to make the rank comparison between them along one of the two dimensions. 
Unknown to the participants, these faces were organized into a 4x4 grid along the two social dimensions; the participants were not instructed on the structure of the grid, and had to infer this structure from trial-and-error learning over pairwise comparisons.

During training, participants only saw trials on which the pair of faces were one rank apart on the appropriate social dimension. 
Then they performed a transitive inference test in the scanner, where they had to make comparisons between faces that were more than one rank apart on the relevant dimension. 
Intriguingly, the researchers found in pilot experiments that participants had trouble learning the task when trials were interleaved, so they implemented a blocked design where participants learned one of the two social dimensions at a time (personal correspondence). 
This is consistent with previous results on similar tasks showing that in some cases learning is improved when trials are blocked \cite{FleschBalaguerDekkerEtAl18a}.  
This task allowed us to explore the learning dynamics in our computational models, but because it was also designed to investigate cognitive map formation in the brain, we were also able to make concrete predictions about the representations that would be learned under different conditions. 

We trained and tested neural network models on the same task structure, including its 4x4 grid and transitive inference test.
However, we introduced two training conditions to the task in order to understand the learning behavior of our models when trials cuing the two dimensions of the grid were blocked or interleaved across training (see Figure \ref{fig:task}).
In the interleaved condition, popularity and competence trials were shuffled randomly, but in the blocked condition the models were trained on one of the two dimensions at a time. 
This allowed us to investigate the potential for cognitive control and gating mechanisms to alleviate the effects of catastrophic forgetting, as has been observed in humans learning certain tasks \cite{CarvalhoGoldstone14, FleschBalaguerDekkerEtAl18a, NohYanBjorkEtAl16, WulfShea02}.

\section{Neural Network Model}
We designed a neural network that leveraged the principles of cognitive control in the PFC, including active maintenance and selective modulation according to the current context or goal. 
To test our hypotheses, we implemented models 1) with and without PFC gating, 2) with different levels of bias to maintain activity over time, and 3) with different levels of control strength. 

\subsubsection{Base Model}
\begin{figure}[ht]
\begin{center}
  \includegraphics[width=0.35\textwidth]{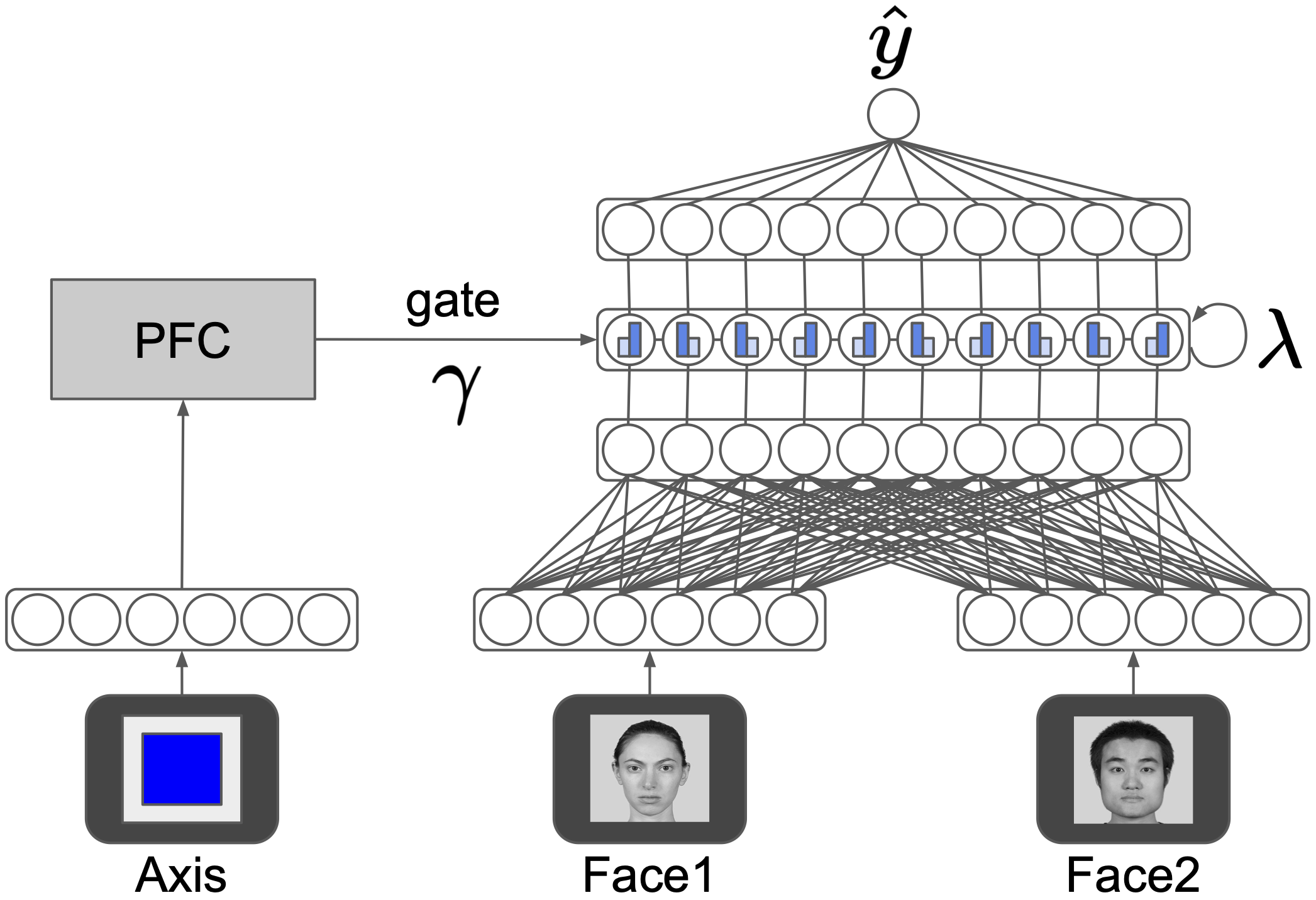}
  \caption{Model architecture. The model was trained to predict which of two faces ranked higher on the cued social dimension (``Axis''). These were given as embeddings, which were passed through an MLP. The activities of units in the hidden layer were modulated by a PFC module, which could gate them via element-wise multiplication by numbers from 0 to 1 (shown as binary probabilities). Additional parameters $\gamma$ and $\lambda$ determined the strength of the control signal and the active maintenance, respectively.}
\label{fig:model}
\end{center}
\end{figure}

To start, we built a simple base neural network with a multi-layer perceptron (MLP) for learning the relationships between the faces in the task (see Figure \ref{fig:model}).
The base model takes three one-hot vectors representing the context cue (``Axis'') and each of the two faces (``Face1'' and ``Face2'') as inputs, and returns a prediction for which face ranked higher on the appropriate dimension.
Each of these three inputs were embedded with linear layers, concatenated, and fed into an MLP with one hidden layer: 
\begin{equation}
    e_a = W_a x_a + b_a \quad e_1 = W_1 x_1 + b_1 \quad e_2 = W_2 x_2 + b_2
\end{equation}
\begin{equation}
    h = \text{ReLU}(W_h [e_a e_1 e_2] + b_h)
\end{equation}
\begin{equation}
\label{eq:output}
    \hat{y} = W_y h + b_y 
\end{equation}
where $x_a$, $x_1$, $x_2$ and $e_a$, $e_1$, $e_2$ are the one-hot vectors and embeddings representing the axis cue, face 1, and face 2, respectively, $h$ is the hidden representation of the MLP, and $\hat{y}$ is the output.
Brackets denote concatenation, and ReLU() is the rectified linear unit activation function. 

\subsubsection{Prefrontal Cortex for Cognitive Control}
In further simulations the base MLP was augmented with a PFC layer that received the context as input and controlled the hidden layer of the MLP by gating its units with an element-wise multiplication:
\begin{equation}
\label{eq:control}
    g = c \odot h 
\end{equation}
where $c$ is a control signal vector generated from the axis cue, and $\odot$ signifies element-wise multiplication. 
The output layer of the MLP then acted on the gated hidden layer, rather than the hidden layer itself (replacing equation \ref{eq:output} above):
\begin{equation}
    \hat{y} = W_y g + b_y
\end{equation}

Note that the PFC layer did not contribute to the output of the model as a whole except through its gating effect on the hidden layer of the MLP. 
This is consistent with classic neural network models of cognitive control \cite{CohenDunbarMcClelland90, MillerCohen01, RougierNoelleBraverEtAl05}, which emphasize the role of the PFC in modulating and regulating the flow of activity in posterior areas through top-down attentional control according to the current goal. 

The control signal was determined from the axis cue according to a simple scheme: half of the units in the hidden layer were gated in response to one of the cues, and the other half of the units were gated in response to the other cue.
\begin{equation}
    c = \begin{cases}
        [1 1 ... 1 0 0 ... 0] \cdot \gamma & \text{if axis} = 0  \\
        [0 0 ... 0 1 1 ... 1] \cdot \gamma & \text{if axis} = 1  
        \end{cases}
\end{equation}
where $\gamma$ determines the strength of the control signal's influence on the hidden units.
Note that there was no learning in the PFC: in this work we were interested in the effects of cognitive control and gating on learning in the MLP when trials were blocked or interleaved. 
Future work will explore methods for introducing learning into the PFC \cite{TsudaTyeSiegelmannEtAl20, WangKurth-NelsonKumaranEtAl18a}, and investigate whether the model could discover a similar scheme.

\vspace{-0.5cm}
\subsubsection{Active Maintenance}
We also implemented a parameter $\lambda$ that controlled a default bias in the PFC layer to maintain its activity over time:
\begin{equation}
\label{eq:lambda}
    s^{(t)} = \sigma(c^{(t)} + \sigma(s^{(t-1)} - 1 + \lambda))
\end{equation}
where $t$ indicates time, $\lambda$ determines the degree to which the previous control signal is added to the current one on each time step, $\sigma$ is a rectified linear function that ensures that the values of the control signal will be between 0 and 1, and now the new variable $s$ integrates the control signal over time and acts on the hidden state of the MLP (replacing equation \ref{eq:control} above):
\begin{equation}
\label{eq:control2}
    g^{(t)} = s^{(t)} \odot h^{(t)}
\end{equation}
The maintenance parameter ($\lambda$) allowed us to control the degree to which the control signal was biased to maintain its activity over time, which introduces a cost when the context (i.e., the axis cue) was switched from trial to trial due to interference from the previous control signal ($s^{(t-1)}$). 
The bias to actively maintain patterns of activity in PFC is well established \cite{OReillyFrank06}, and is fundamental to the important role the PFC plays in working memory, executive functioning, and planning.
We hypothesized that these dynamics would be relevant to our setting because when trials are interleaved the switch cost may have negative effects on learning. 
We used a particularly simple implementation to capture this basic dynamic, but future work will investigate whether its effects on learning play out in more realistic implementations \cite{OReillyFrank06}.

\begin{figure*}[ht]
\begin{center}
  \includegraphics[width=0.95\textwidth]{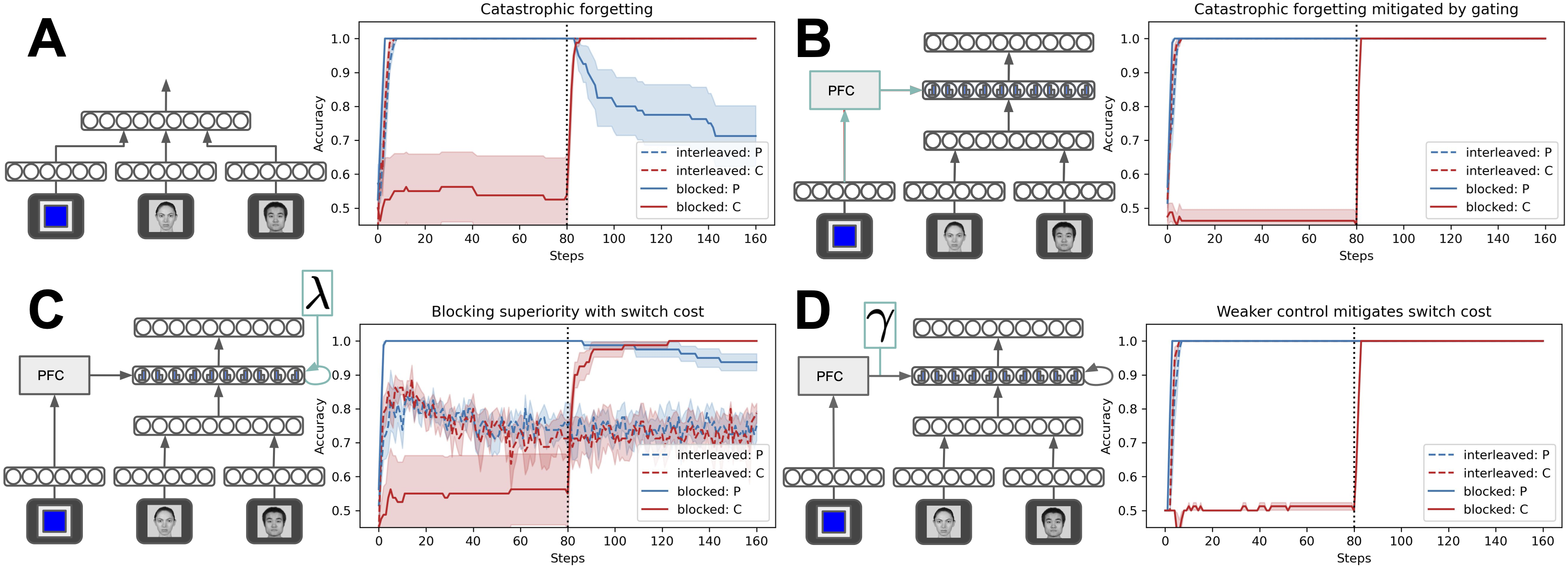}
  \caption{Accuracy results. Each plot shows accuracy (y-axis) over the course of training steps (x-axis) for a configuration of the model, depicted by a diagram next to the associated plot. In each experiment, trials were split on the test set by the relevant axis cue (popularity = P, shown in blue, and competence = C, shown in red), and accuracy was measured separately for each in order to show the effects of blocking. Each simulation included 5 runs where trials were interleaved (dashed lines) and 5 runs where trials were blocked (solid lines). Solid areas show SEM across runs. \textbf{A)} Catastrophic forgetting occurred in the base MLP model when it was trained on the blocked condition. \textbf{B)} Catastrophic forgetting was alleviated by the addition of a control signal from the PFC module (highlighted in cyan). \textbf{C)} The model's performance on the interleaved condition suffered when a default active maintenance was introduced in the control signal (shown by self-connection highlighted in cyan), inducing a cost to switching between contexts. \textbf{D)} This switch cost was eliminated when the control strength was reduced (shown by $\gamma$ highlighted in cyan), demonstrating the tradeoff between control strength and switch cost.}
\label{fig:results}
\end{center}
\end{figure*}

\subsubsection{Implementation Details}
Models were built in PyTorch, and were supervised on correct responses with a cross entropy loss function.
Models were optimized using backpropagation and Adam \cite{KingmaBa15} with a learning rate of 0.001. 
Embedding vectors had 32 dimensions, and there were 128 units in the hidden layer.
For each simulation, 5 runs with different random initializations were performed.

\section{Results}
All versions of the model were trained on both blocked and interleaved conditions. 
In particular, we explored our hypotheses by testing the model with different configurations of the parameters described above.
Accuracy on the test set was evaluated for each social dimension separately in order to assess forgetting in the blocked condition.

\subsection{Catastrophic Forgetting when Trials are Blocked}
First, we reproduced catastrophic forgetting in the model by training the base MLP (without a PFC) on both the blocked and interleaved conditions of the task (see Figure \ref{fig:results}A). 
When trials were interleaved, the base MLP model had no problem learning the task, and quickly achieved 100\% accuracy on the test set. 
However, when trials were blocked, we observed catastrophic forgetting: after initially performing well on the first block, over the course of the second block performance progressively declined, indicating increasing forgetting of the relationships along the first dimension that were learned in the preceding block.
This result can be understood in the context of CLS theory \cite{McClellandMcNaughtonOReilly95}, which suggests that catastrophic forgetting occurs whenever overlapping patterns interfere with each other. 

\subsection{Cognitive Control Mitigates Forgetting}
To establish that gating in the PFC can mitigate interference and reduce catastrophic forgetting, we trained the model equipped with a PFC on the same set of conditions (see Figure \ref{fig:results}B). 
For the purposes of this experiment, we removed the internal dynamics of the PFC, setting the $\lambda$ parameter to 0 (no maintenance) and the $\gamma$ parameter to 1.0. 
When this model was trained on the task, its performance on interleaved trials was unaffected, and quickly rose to 100\% accuracy.
However, when it was trained on blocked trials, the catastrophic forgetting observed in the previous experiment was alleviated, and the model was capable of retaining what it had learned in the first block through the subsequent block. 

This finding is consistent with the basic principles of CLS \cite{McClellandMcNaughtonOReilly95}: when the overlap between patterns of activity in the hidden layer is reduced, interference and forgetting are alleviated.
However, CLS theory holds that the hippocampus reduces overlap in its representations with mechanisms that promote sparsity, whereas here we show that a PFC equipped with a dynamic gating mechanism can accomplish a similar goal.
This is consistent with the results of previous computational models \cite{RougierNoelleBraverEtAl05, TsudaTyeSiegelmannEtAl20} showing that adaptive gating can offer an alternative mechanism for reducing the overlap between patterns of activity, thereby reducing interference and forgetting.

\subsection{Blocking Advantage with a Switch Cost}
The results above and the results of previous models \cite{RougierNoelleBraverEtAl05, TsudaTyeSiegelmannEtAl20} show that catastrophic forgetting can be reduced when learning occurs in non-overlapping patterns of activity across a layer, thereby explaining the reduced effects of interference observed in humans and other animals as compared with standard neural network models.
However, in certain cases human performance has been shown to be \textit{superior} when trials are blocked compared with when they are interleaved \cite{CarvalhoGoldstone14, FleschBalaguerDekkerEtAl18a, NohYanBjorkEtAl16}.
We hypothesized that this reversal of the catastrophic forgetting phenomenon may be due to the internal dynamics of cognitive control processes, and in particular due to the bias in neurons in the PFC to actively maintain their activity over time \cite{OReillyFrank06}.
To explore this hypothesis, we implemented a control model with simple recurrent dynamics (see Equation \ref{eq:lambda}), keeping the $\gamma$ parameter at 1.0 but setting the $\lambda$ parameter to 0.9 (i.e., 90\% of the previous control signal is maintained at each time step). 
The resultant dynamics can be thought of as exhibiting a switch cost \cite{BlackwellChathamWiseheartEtAl14, HyafilSummerfieldKoechlin09}, wherein rapidly switching the context or goal (in this case the relevant social dimension) introduces interference due to the ongoing maintenance of the previous context.
Note that the cognitive cost of task switching is usually measured in increased reaction times or errors, but here we study it in the context of its effects on learning. 

\begin{figure}[ht]
\begin{center}
  \includegraphics[width=0.35\textwidth]{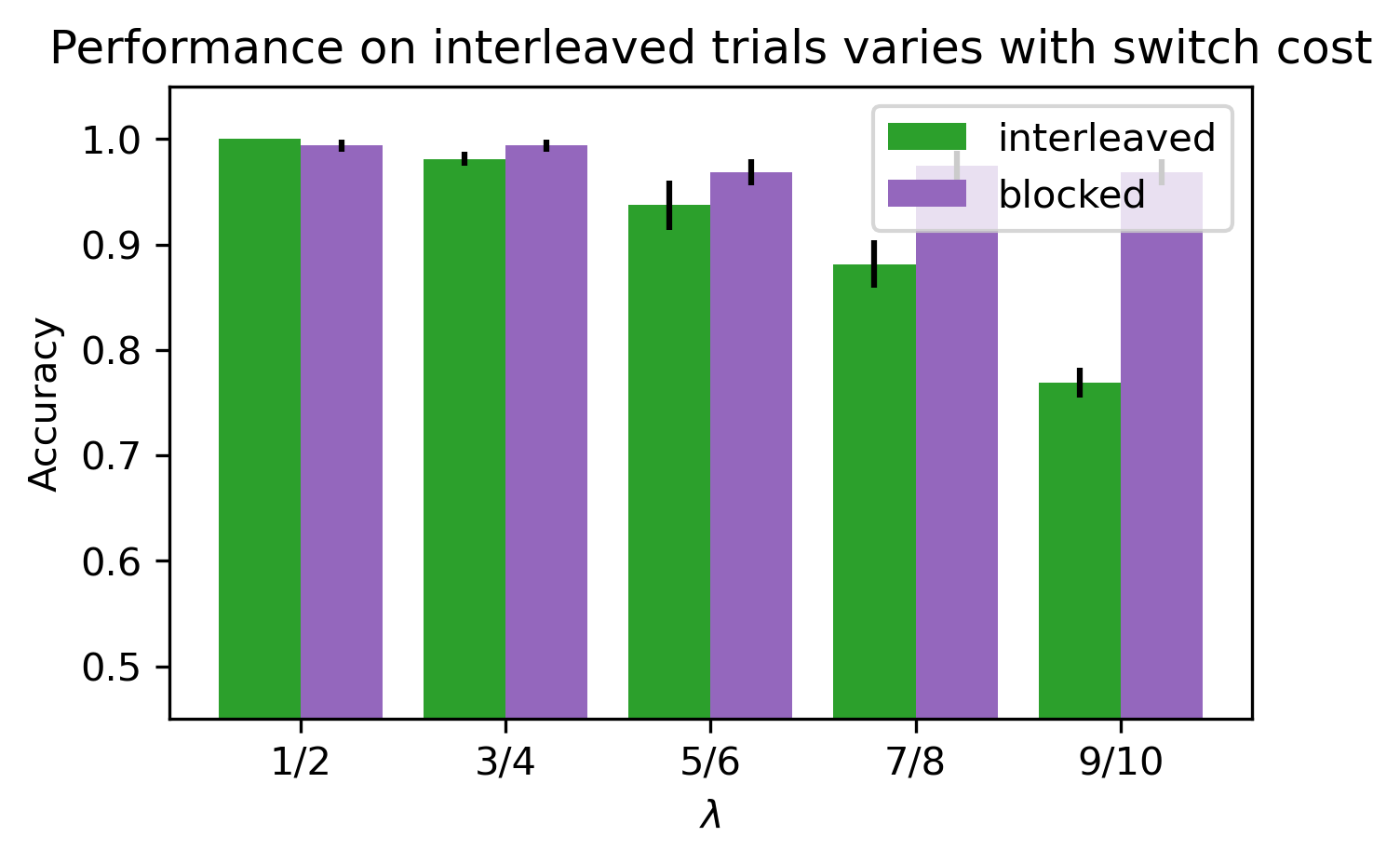}
  \caption{Effect of maintenance parameter ($\lambda$) on performance. In the blocked condition (purple), accuracy on the test set does not depend much on maintenance. However, as maintenance increases, the cost to switching worsens and performance in the interleaved condition declines.}
\label{fig:maintenance}
\end{center}
\end{figure}

When these dynamics were introduced, the model was relatively unaffected when trials were blocked, but exhibited a consistent reduction in performance when trials were interleaved (see Figure \ref{fig:results}C). 
When trials were interleaved, many switches between contexts occurred throughout training, thereby introducing interference in the control signal, causing processing to be ineffectively modulated according to the current context.
We also performed simulations where we systematically varied the $\lambda$ parameter (see Figure \ref{fig:maintenance}), showing consistent reductions in performance on the interleaved condition with increased active maintenance.

\subsection{Tradeoff between Control Strength and Switch Cost} 
Previous work has suggested a natural tradeoff between the strength of cognitive control and the cost incurred when a context or task-set is switched \cite{HerdOReillyHazyEtAl14}: stronger control would be more effective in coordinating activity in other brain regions according to the current goal, but may make rapid switching between task sets or goals more difficult.
To demonstrate this tradeoff, we tested a model with the maintenance ($\lambda$) kept at 0.9, but reduced the value of $\gamma$ (control strength) to 0.1. 
In this case, the model still performed well when trials were blocked, but the reductions in performance when trials were interleaved disappeared (see Figure \ref{fig:results}D).
This shows that weakening the control signal can reduce the switch cost, aiding performance when there are many switches. 
Our results are consistent with a tradeoff between the strength of control and the switch cost: without control, catastrophic forgetting is detrimental to performance when trials are blocked, but when control is too strong, interference hurts performance when trials are interleaved.

\subsection{Analysis of Learned Representations}

\begin{figure*}[ht]
\begin{center}
  \includegraphics[width=0.9\textwidth]{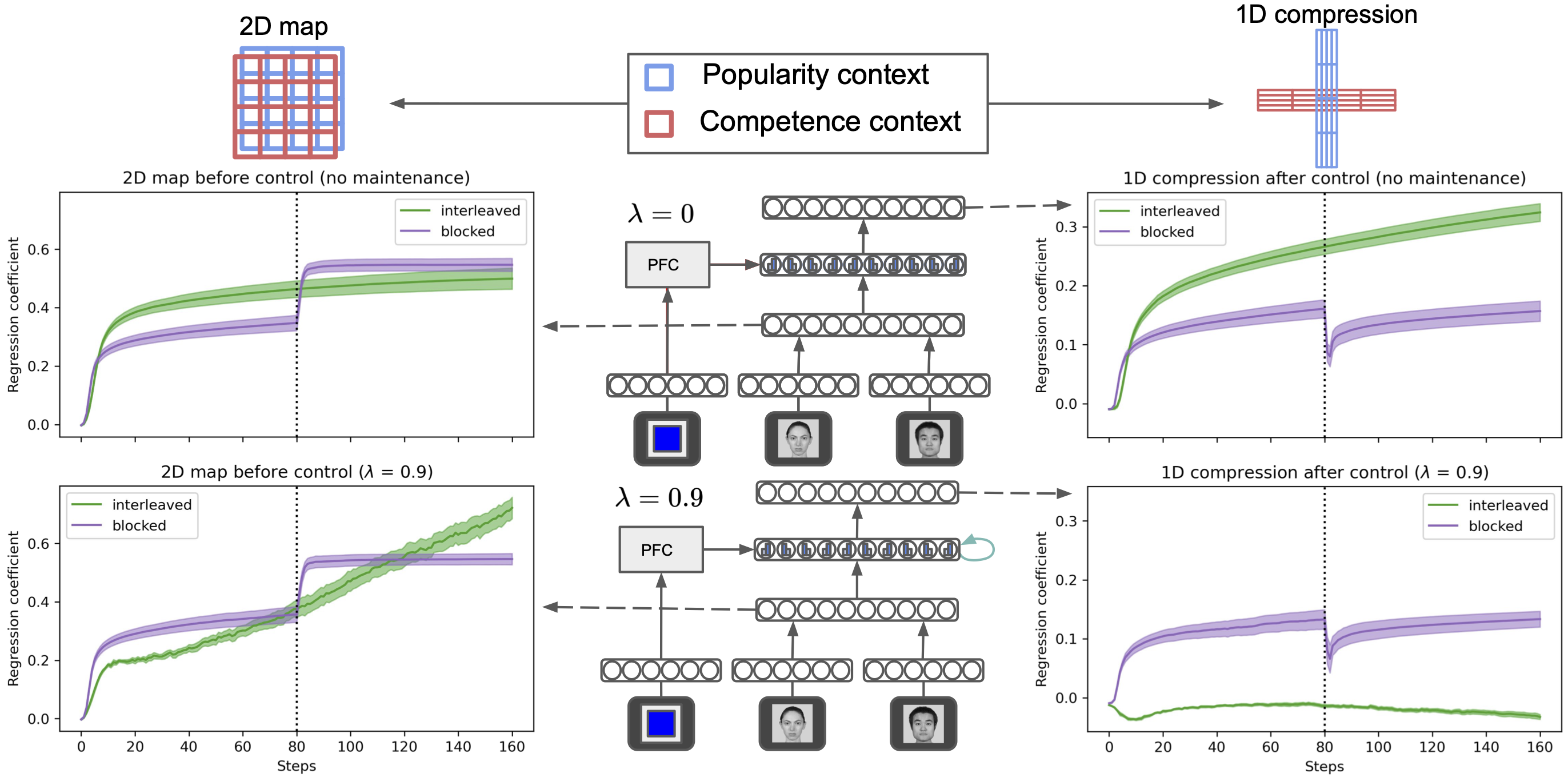}
  \caption{Results of analyzing the learned representations of the model. Representations were analyzed in terms of how well they captured the 2D structure of the 4x4 grid (left) and how much the irrelevant dimension of the grid was compressed on each trial (right). Idealized grids depicting these two predictions are shown on the top, where the red grid indicates idealized spacing between representations extracted during trials on which competence was cued, and the blue grid indicates the same for popularity trials. Plots show the beta coefficients over training from performing the relevant regressions. These were conducted on hidden representations either before (left) or after (right) control was applied, on two configurations of the model - one with $\lambda = 0$ (no maintenance) and one with $\lambda = 0.9$. Regression results revealed strong 2D map-like structure in the hidden layer before control was applied, and strong 1D compression of the irrelevant dimension after control was applied. However, when the active maintenance was too strong ($\lambda = 0.9$), the compression effect disappeared in the interleaved condition, indicating a failure to modulate representations according to the current context. Vertical lines indicate the switch in the blocked condition.}
\label{fig:analyses}
\end{center}
\end{figure*}

The grid structure of our task allowed us to make concrete predictions about the representations that would be learned in the hidden layers of the network \cite{ParkMillerNiliEtAl20, RussinZolfagharParkEtAl21}.
In particular, we tested whether the model formed 2D map-like representations that captured the basic structure of the grid \cite{ConstantinescuOReillyBehrens16, ParkMillerNiliEtAl20, ParkMillerBoorman21a}, and whether these 2D map-like representations were modulated by the current context.
Previous work has shown that on a similar task, 2D structure was modulated by the current context, compressing the irrelevant dimension \cite{FleschJuechemsDumbalskaEtAl21}.

Figure \ref{fig:analyses} shows the results of performing a regression on the representations from the hidden layer with hypothetical distance matrices (depicted as idealized map-like representations) as the predictors.
We compared the results of this regression throughout training when the maintenance parameter ($\lambda$) was set to 0 and 0.9, and when the hidden representations were extracted before and after the control signal was applied (see equation \ref{eq:control2}).  

The model reliably learned the 2D structure of the grid in its hidden representations regardless of the maintenance, as can be seen in the results from the hidden representations before the control signal was applied. 
This 2D structure was modulated by the current control signal, which had the effect of compressing the currently irrelevant dimension (or equivalently, expanding the relevant dimension).
This suggests that the effect of the control signal was to allow the model to generate its response based on the relevant dimension, and to appropriately facilitate learning in the neurons coding for that dimension. 
However, when trials were interleaved and maintenance ($\lambda$) was set to 0.9, the model did not show this compression pattern after control was applied, indicating a failure to modulate its representations according to the current context. 
This confirmed the idea that the poor performance on interleaved trials when the switch cost was high (see Figure \ref{fig:results}C) was caused by interference in the control signal. 

\section{Discussion}

The neural networks driving current ML research do not perform well in continual-learning settings where incoming data is blocked or otherwise correlated over time \cite{HadsellRaoRusuEtAl20}.
Humans do not exhibit the catastrophic forgetting that plagues these neural networks in these settings \cite{McClellandMcNaughtonOReilly95}, and in some cases even show a learning advantage when trials are blocked \cite{CarvalhoGoldstone14, FleschBalaguerDekkerEtAl18a}.
In this work, we built on previous computational frameworks \cite{FleschBalaguerDekkerEtAl18a, RougierNoelleBraverEtAl05, TsudaTyeSiegelmannEtAl20}, and investigated the potential for cognitive control mechanisms in the PFC to induce non-overlapping patterns of activity in order to mitigate interference. 
Consistent with previous studies \cite{TsudaTyeSiegelmannEtAl20}, our simulations suggest that these mechanisms can aid learning when trials are blocked over time.

In addition to pattern-separation mechanisms in the hippocampus proposed in CLS \cite{McClellandMcNaughtonOReilly95}, and the gating mechanism in PFC proposed here and elsewhere \cite{RougierNoelleBraverEtAl05, TsudaTyeSiegelmannEtAl20} a number of alternative mechanisms for alleviating catastrophic forgetting in neural networks have been explored \cite<e.g., >{FleschBalaguerDekkerEtAl18a, KirkpatrickPascanuRabinowitzEtAl17a, VelezClune17a}.
In particular, \citeA{FleschBalaguerDekkerEtAl18a} show that forgetting was reduced on a similar task when their network was augmented with a good inductive prior.
However, they did not show an advantage to blocking over interleaving, although they observed this effect in their human experiments. 
While our approach is not incompatible with the idea that good inductive priors can mitigate catastrophic forgetting, we also show that a bias to maintain activity in the PFC leads to an advantage of blocking over interleaving, providing an explanation for some of the results observed by \citeA{FleschBalaguerDekkerEtAl18a} and others.

The advantage of blocking over interleaving observed in human learning can seem to contradict the well-established principles of CLS \cite{McClellandMcNaughtonOReilly95}. 
However, we show here that a neural system equipped with mechanisms for cognitive control and active maintenance can enter a different regime than those typically considered in the CLS framework, wherein a reliance on control exposes the system to interference in the control signal caused by rapid context switches. 
We speculate that in most cases, pattern-separation mechanisms in the hippocampus are sufficient to ensure effective learning regardless of whether experiences are correlated over time, but animals such as humans that rely heavily on cognitive control may in some cases \textit{require} learning experiences to be correlated over time due to the bias for active maintenance in the cognitive controller.
In our simulations, we introduced this bias to show how it could lead to a learning advantage of blocking, but of course there was no real need for active maintenance in the task (as shown by the good performance of the base MLP when trials were interleaved). 
We expect that there are good computational reasons that the PFC would have a bias to maintain its activity over time (e.g., related to its role in working memory and planning), and that these may be unrelated to the demands of this particular task.
We leave it to future work to show that a system augmented with cognitive control and active maintenance is superior in an absolute sense to one without these mechanisms.   

Our simulations were also inspired by the idea that active maintenance engenders a cost to switching between contexts, which must be traded off against the strength with which control can be applied \cite{HerdOReillyHazyEtAl14}.
The presence of this tradeoff means the cognitive system as a whole must optimize the strength of its control signal according to constraints imposed by learning as well as the current need for control \cite{ShenhavBotvinickCohen13a}.
This optimization may have taken place over the course of evolution \cite{HerdOReillyHazyEtAl14}, but it may also occur in real time according to the task at hand \cite{OReillyNairRussinEtAl20}. 

Representational analyses showed that cognitive control can act on 2D map-like representations to modulate them according to the current context by compressing irrelevant dimensions and allowing learning to take place in non-overlapping patterns. 
However, a strong bias to maintain activity over time leads to interference in the control signal, reducing this effect and leading to poor performance when trials are interleaved.
\citeA{FleschJuechemsDumbalskaEtAl21} also showed compression along currently irrelevant dimensions in representations of a neural network trained on a similar task.
In particular, this occurred in a ``rich'' regime when their neural network was initialized with small weights.
The default random initializations we used in our model were likely small enough to put them in the ``rich'' regime, but future work will assess the extent to which our results depend on initialization.

Intelligent systems should be capable of continually learning in settings where data is not independently sampled over time. 
Our simulations demonstrate computational principles that may underlie human continual learning, and help to explain behavioral phenomena observed in human experiments.

\section{Acknowledgments}
We would like to thank the members of the Computational Cognitive Neuroscience lab and the Learning and Decision Making lab, as well as reviewers for helpful comments and discussions. 
The work was supported by: ONR grants ONR N00014-20-1-2578, N00014-19-1-2684 / N00014-18-1-2116, N00014-18-C-2067, as well as NSF CAREER Award 1846578, and NIH R56 MH119116.

\bibliographystyle{apacite}

\setlength{\bibleftmargin}{.125in}
\setlength{\bibindent}{-\bibleftmargin}

\bibliography{main}

\end{document}